\documentclass[11pt, preprint]{article}
\usepackage{color} \usepackage{cancel}
\usepackage{float} \usepackage{graphicx}% Include figure files
\usepackage{subfigure} \usepackage{caption}	% Advanced figure caption
\usepackage{dcolumn}% Align table columns on decimal point
\usepackage[toc,page]{appendix} \usepackage{authblk}	 		% Title page
\usepackage{indentfirst}	% Indent for first paragraph
\usepackage{authblk}	 		% Title page
\usepackage{multirow}     % Multi-row cell in table
\usepackage{hyperref}     % Internal and external hyper links
\usepackage{amsfonts, amssymb, amsmath}
\usepackage[left=2.5cm,right=2.5cm,top=2.9cm,bottom=2.9cm,a4paper]{geometry}
\usepackage[noadjust]{cite}
\usepackage{filecontents}
\providecommand{\keywords}[1]{{{Keywords:}} #1}

\usepackage{soul}
\usepackage{xcolor}
\setstcolor{red}
\graphicspath{{./figures/}}
\hypersetup{ colorlinks   = true, % Color links instead of ugly boxes
urlcolor     = blue, % Color of external hyperlinks
linkcolor    = blue, % Color of internal links
citecolor    = red 	 % Color of citations }
}

\title{\Large\bf Exact and slow-roll solutions for  exponential power-law inflation connected with f(R) gravity and observational constraints}
\author[1]{I. V. Fomin\thanks{ingvor@inbox.ru}}
\author[1,2,3]{S. V. Chervon\thanks{chervon.sergey@gmail.com}}
\affil[1]{\small \it  Bauman Moscow State Technical University, 2-nd Baumanskaya street, 5, Moscow, 105005, Russia}
\affil[2]{\small \it Ulyanovsk State Pedagogical University, Lenin's Square, 4/5, 432071, Ulyanovsk,
Russia}
\affil[3]{\small \it Kazan Federal University, Kremlevskaya street 18, Kazan, 420008, Russia}

\begin{document}

\maketitle
%----------------------------------------------------------------
\begin{abstract}
We investigate an ability of the exponential power-law inflation to be
phenomenologically correct model of the early universe.
GR scalar cosmology equations we study in Ivanov-Salopek-Bond (or Hamilton-Jacobi like) representation where the Hubble parameter $H$ is the function of a scalar field $\phi$.
Such approach admits calculation of the potential for given $H(\phi)$ and consequently reconstruction of $f(R)$ gravity in parametric form.
By this manner the Starobinsky potential and non-minimal Higgs potential (and consequently the corresponding $f(R)$ gravity) were reconstructed using constraints on model's parameters.
Also comparison to observation (PLANCK 2018) data shows that both models give correct values for scalar spectral index and tensor-to-scalar ratio under wide range of exponential-power-law model's parameters.
%
%It is shown that the Starobinsky and non-minimal Higgs inflation correspond to this model in Einstein frame.

%Also, we consider the verification of given inflationary scenario on the basis of the parameters of cosmological perturbations.
\end{abstract}
%\tableofcontents


%----------------------Introduction------------------------------
\section{Introduction}

The inflationary paradigm implying the accelerated expansion of the early universe is often considered as a successful
explanation for the origin of it's structure. The first models of cosmological inflation were built on the basis of General Relativity
(GR) in 4D Friedmann-Robertson-Walker (FRW) space-time under the assumption of the existence of some scalar field
(inflaton) which is the source of the accelerated expansion of the
universe~\cite{Starobinsky:1980te,Guth:1980zm,Linde:1981mu,Linde:2005ht,Lyth:2009zz}.

Since the potential of a scalar field $V(\phi)$ has a key role in the construction of inflation scenarios, the models of the early
universe are often determined by the chosen potential. Such a method  can be called as the ``potential motivated approach".
In this case, inflationary models are classified by the scalar field potential.

On the other hand, the dynamics of the expansion which is characterized by a scale factor $a(t)$ no less important for
understanding
of the inflationary scenarios and one can reconstruct the potential $V(\phi)$ from the chosen scale
factor (see, for example, \cite{Zhuravlev:1998ff}).
In the context of this method, which we will call the ``dynamical motivated approach",  inflationary models are classified by the
laws of the expansion of the universe.

A convenient tool for the analysis of the inflationary model  is the slow-roll
approximation~\cite{Liddle:1994dx,Lyth:2009zz,Linde:2005ht} and a lot of models are considered on the basis of this
approach. Nevertheless, for a more correct understanding of the nature of processes at the stage of inflation, the exact solutions
of cosmological dynamic equations were considered as well (see, for example, \cite{Chervon:2017kgn,Fomin:2017xlx}).
The classification of the methods for generating them (and the exact solutions themselves) for inflation based on the Einstein
gravity can be found  in the monograph~\cite{Chervon:2019sey}.

The inflationary models containing a combination of Friedmann solutions and (quasi) de Sitter solutions constitute the basis of an actual description of the evolution of the early universe. In the context of the inflationary paradigm, the early universe expands rapidly for some time and then goes into a power-law expansion regime without acceleration corresponding to the Friedmann solutions for radiation and barionic matter.

In this paper we consider the exponential power-law inflation which implies such a combination of the dynamical regimes on the basis of the exact solutions and ones obtained in slow-roll approximation and show that widely discussed Starobinsky~\cite{Starobinsky:1980te,Whitt:1984pd,Maeda:1987xf,Ketov:2014jta,Motohashi:2017vdc,Aldabergenov:2018qhs} and non-minimal Higgs~\cite{Bezrukov:2007ep,Bezrukov:2008ej,Gorbunov:2014ewa} inflationary models which imply the same potential of the scalar field can be considered as the partial cases of the exponential power-law inflation in the context of the dynamical motivated approach.

The article is organising as follow. Sec.2 contains exact and slow-roll GR cosmology equations in FRW spacetime. Also the relations a scalar field gravity to f(R) gravity are reproduced from the work~\cite{Aldabergenov:2018qhs}. Sec. 3 is devoted to analysis of the slow-roll solutions
for exponential power-law (EPL) inflation. Three parametric slow-roll approximated solution was found and it was shown which restrictions on parameters lead to Starobinsky potential (which, in turn, leads to Starobinsky $R+R^2$ gravity model). Analysis of exact EPL inflation is performed in Sec. 4. There were shown the derivation of de Sitter solution, Starobinsky gravity using special choice of EPL model parameters. The relation between exact and slow-roll solutions is discussed as well.
In Sec. 5 it was considered correspondence EPL inflation to observation data. It was shown that one of the parameters can always be chosen so that observational constraints are satisfied. In Sec. 6 we list the main results of investigation.

\section{The exact and approximate solutions in cosmology and conformal connection with \texorpdfstring{$f(R)$}{fR}-gravity}\label{section2}

The inflationary models based on Einstein gravity and a single scalar field defined by the action
\begin{equation}
\label{E}
S=\int d^{4}x\sqrt{-g}\left[\frac{1}{2}R-\frac{1}{2}g^{\mu\nu}\partial_{\mu}\phi\partial_{\nu}\phi -V(\phi)\right],
\end{equation}
where $\phi$ is a scalar field, $V(\phi)$ is the potential of a scalar field and $g^{\mu\nu}$ is a metric tensor of a space-time. We set Einstein gravitational constant $\varkappa=8\pi G=1$.

The variation of the action (\ref{E}) with respect to the metric and field  in a spatially flat
Friedmann-Robertson-Walker space
\begin{equation}
ds^2=-dt^2+a^{2}(t)\left(dx^{2}+dy^{2}+dz^{2}\right),
\end{equation}
gives three dynamic equations
\begin{equation}
 \label{DE1}
3H^{2}=\frac{1}{2}\dot{\phi}^{2}+V(\phi)\equiv\rho_{\phi},
\end{equation}
\begin{equation}
\label{DE2}
-3H^{2}-2\dot{H}=\frac{1}{2}\dot{\phi}^{2}-V(\phi)\equiv p_{\phi},
\end{equation}
\begin{equation}
\label{DE3}
 \ddot{\phi} + 3H\dot{\phi} +V'_{\phi}= 0,
\end{equation}
where $\rho_{\phi}$ and $p_{\phi}$ are the energy density and the pressure of a scalar field, also, $V'_{\phi}=dV/d\phi$.

The methods of the exact solutions construction for these system of equations one can find, for example,
in~\cite{Fomin:2017xlx,Chervon:2017kgn}. One of them is the method proposed by Ivanov~\cite{givanov81} and subsequently
Salopek and Bond~\cite{Salopek:1990re}.

From the equations (\ref{DE1})--(\ref{DE3}) only two are independent, and this system can be represented as the
Ivanov-Salopek-Bond equations (or as Hamilton-Jacobi type equations)
\begin{eqnarray}
\label{ISB1}
&&V(\phi)=3H^{2}-2H'^{2}_{\phi},\\
\label{ISB2}
&&\dot{\phi}=-2H'_{\phi},
\end{eqnarray}
in which the exact solutions are obtained by the choice of the Hubble parameter $H(\phi)$.

In the case of the slow-roll approximation which implies that $V(\phi)\gg\frac{1}{2}\dot{\phi}^{2}$ and
$\ddot{\phi}\approx0$ the system (\ref{DE1})--(\ref{DE3}) is reduced to the equations
 \begin{eqnarray}
\label{ISB1SB}
&&V(\phi)\approx3H^{2},\\
\label{ISB2SB}
&&\dot{\phi}\approx-2H'_{\phi}.
\end{eqnarray}

Therefore, the difference between the exact and approximate background cosmological solutions is the second term in the
potential (\ref{ISB1}).

To make comparison of the scalar field gravity \eqref{E} with $f(R)$-gravity with the action
\begin{equation}
S=\int d^4x \sqrt{-g}\left[f(R)\right],
\end{equation}
we will use the following relations~\cite{Aldabergenov:2018qhs}
\begin{eqnarray}
\label{R}
R = \left[ \sqrt{6}\,\frac{dV}{d\phi} + 4V\right] \exp \left( \sqrt{\frac{2}{3}}\phi \right), \\
\label{F}
f= \left[ \sqrt{6}\,\frac{dV}{d\phi} + 2V\right] \exp \left( 2\sqrt{\frac{2}{3}}\phi \right),
\end{eqnarray}
which connect $f(R)$-gravity and models based on Einstein gravity in parametric form.
Thus, one can use the relation \eqref{R}-\eqref{F} to put in accordance cosmological models based on Einstein gravity and $f(R)$-gravity on the basis of the exact (\ref{ISB1}) and approximate (\ref{ISB1SB}) expressions for the potential of a scalar field.

\section{The slow-roll solutions for exponential power-law inflation}\label{section3}

%Now, we consider the model with the Hubble parameter
We represent the exponential power-law (EPL) expansion of universe, with following parametrisation of the Hubble function
\begin{equation}
\label{Hubble}
H(\phi)=-\mu_{1}\exp(-\mu_{2}\phi)+\mu_{3},
\end{equation}
where $\mu_{1}$, $\mu_{2}$ and $\mu_{3}$ are an arbitrary constants.

The solutions of the equations (\ref{ISB1SB})--(\ref{ISB2SB}) in slow-roll approximation are:
\begin{eqnarray}
\label{POTEPLSR-1}
&&V_{SR}(\phi)=3\mu^{2}_{1}\exp(-2\mu_{2}\phi)-6\mu_{1}\mu_{3}\exp(-\mu_{2}\phi)+3\mu^{2}_{3},\\
\label{field-1}
&&\phi(t)=\frac{1}{\mu_{2}}\ln\left(-2\mu_{1}\mu^{2}_{2}t-c\right),\\
\label{HubbleSR-1}
&&H(t)=\frac{\mu_{1}}{2\mu_{1}\mu^{2}_{2}t+c}+\mu_{3},\\
\label{scalefactorSR-1}
&&a(t)=a_{0}\exp(\mu_{3}t)(2\mu_{1}\mu^{2}_{2}t+c)^{1/2\mu^{2}_{2}},
\end{eqnarray}
where $c$ is the constant of integration.

For the case $2\mu_{1}\mu^{2}_{2}t+c>0$, the scalar field (\ref{field-1}) can be considered as a complex one with the variable real part and constant imaginary part \cite{Abramowitz}
\begin{equation}
\label{CSF}
\phi(t)=\frac{1}{\mu_{2}}\ln\left(2\mu_{1}\mu^{2}_{2}t+c\right)+\frac{i\pi}{\mu_{2}},
\end{equation}
and for $2\mu_{1}\mu^{2}_{2}t+c<0$  the scalar field (\ref{field-1}) is real
\begin{equation}
\label{CSFcan}
\phi(t)=\frac{1}{\mu_{2}}\ln\left(2\mu_{1}\mu^{2}_{2}t+c\right).
\end{equation}

Also, we note, that the scalar field (\ref{field-1}) has the positive kinetic energy
\begin{equation}
\label{KinE}
X=\frac{1}{2}\dot{\phi}^{2}=\frac{2\mu^{2}_{1}\mu^{2}_{2}}{\left(-2\mu_{1}\mu^{2}_{2}t-c\right)^{2}}\geq0,
\end{equation}
therefore, for both these cases the lagrangian of this field has the canonical form ${\mathcal L}=X-V$.

Looking for connection of the potential (\ref{POTEPLSR-1}) with Starobinsky potential~\cite{Starobinsky:1980te}
we set $\mu_{2}=\sqrt{2/3}$. Then, from the equations (\ref{R})--(\ref{F}) we derive
\begin{equation}
f(R)=\frac{\mu_{1}}{\mu_{3}}R+\frac{1}{24\mu^{2}_{3}}R^{2}.
\end{equation}
Further, choosing the model's parameters
 \begin{equation}
\label{constantsSR}
\mu_{1}=\mu_{3}=\pm\frac{1}{2}m,
\end{equation}
from (\ref{POTEPLSR-1}) we get the potential
\begin{equation}
\label{Starobinsky}
V_{SR}(\phi)=\frac{3}{4}m^{2}\left(1-e^{-\sqrt{\frac{2}{3}}\phi}\right)^{2},
\end{equation}
which exactly corresponds to the Starobinsky potential~\cite{Starobinsky:1980te,Whitt:1984pd,Maeda:1987xf,Aldabergenov:2018qhs} and non-minimal Higgs potential \cite{Bezrukov:2007ep,Bezrukov:2008ej} as well.
The parameter $m=1.13\times 10^{-5}$~\cite{Mishra:2018dtg} can be considered as the mass of the scalar field.
Also, we note, that this potential leads to the Starobinsky gravity~\cite{Aldabergenov:2018qhs}
\begin{eqnarray}
\label{StarobinskyFR}
f(R)=R+\frac{1}{6m^{2}}R^{2},
\end{eqnarray}
which generalized the Einstein gravity by the second quadratic term in curvature.
%%%%%%%%%%%
Starobinsky gravity model as a special case of $f(R)$ gravity was considered in astrophysics and cosmology (see, for example, \cite{1-Savas}, \cite{2-Giudice}, and literature cited therein). Also the relation of Starobinsky model to modified gravity and supergravity is discussed in \cite{3-Channuie} \cite{5-Sebastiani}, \cite{6-Moraes}, \cite{7-Odintsov}, \cite{4-Ketov}.

%%%%%%%%%%%

Substituting the parameters (\ref{constantsSR}) with upper sign into the solutions (\ref{field-1})--(\ref{scalefactorSR-1})
we obtain
\begin{eqnarray}
\label{fieldsr}
&&\phi(t)_{inf}=\sqrt{\frac{3}{2}}\ln\left(-\frac{2}{3}mt-c\right),\\
\label{Hubblesr}
&&H(t)_{inf}=\frac{m}{\frac{4}{3}mt+2c}+\frac{1}{2}m,\\
\label{scalefactorsr}
&&a(t)_{inf}=a_{0}\exp\left(\frac{1}{2}mt\right)(2mt+3c)^{3/4}.
\end{eqnarray}

For the case of lower sign we obtain the solutions
\begin{eqnarray}
\label{fieldsr1}
&&\phi(t)_{def}=\sqrt{\frac{3}{2}}\ln\left(\frac{2}{3}mt-c\right),\\
\label{Hubblesr1}
&&H(t)_{def}=\frac{m}{\frac{4}{3}mt-2c}-\frac{1}{2}m,\\
\label{scalefactorsr1}
&&a(t)_{def}=a_{0}\exp\left(-\frac{1}{2}mt\right)(2mt-3c)^{3/4},
\end{eqnarray}
corresponding to deflationary scenario~\cite{Gasperini:1993hu}.

After the following redefinition of time $t\rightarrow-t$, the scale factor (\ref{scalefactorsr1}) for negative constant $c<0$ is transformed to
\begin{equation}
a(t)=a_{0}\exp\left(\frac{1}{2}mt\right)(-2mt+3c)^{3/4},
\end{equation}
which corresponds to the collapse of the universe at the time $t_{\ast}=3c/2m$.

Therefore, despite the fact that the potential (\ref{Starobinsky}) implies two possible solutions, the inflationary solutions
(\ref{fieldsr})--(\ref{scalefactorsr}) only have a phenomenological correspondence with the evolution of the early universe.

\subsection{The dynamics of the universe's expansion}

Requesting positive sign for $e^{\mu_2\phi}$ (or, equivalently, positive value for logarithm's argument of \eqref{field-1} we have the restriction on time
\begin{equation}\label{rstr-t}
t<t_{end},~~t_{end}=c/(2\mu_1\mu_2^2)
\end{equation}
when $c \mu_1<0$.  It means that till that time early inflation should get finish.
%$c>0,~\mu_1<0$ or $c<0,~\mu_1>0$

Further restriction we obtain from universe expansion, i.e. $H>0$.

From equation \eqref{HubbleSR-1} we can find that $H$ will be positive always if
\begin{equation}\label{H-pos-1}
\mu_3>0,~~\mu_1<0,~~c>0
\end{equation}

If $\mu_3<0$ the first term in rhs of \eqref{HubbleSR-1} should be positive and excess $|\mu_3|$. This leads to restriction on time
\begin{equation}\label{H-pos-2}
t<|\frac{\mu_1/|\mu_3|-c}{2\mu_1\mu_2^2}|
\end{equation}
which is valid for $c>0,~\mu_1<0$ or $c<0,~\mu_1>0$.
So the universe will be under expansion till the time \eqref{H-pos-2} which should be less then validate of the solution, the time $t_{end}$ \eqref{rstr-t}. Therefore the inequality
\begin{equation}
|\mu_3|<\mu_1/(2c)
\end{equation}
should be true.

Let us study the period of acceleration for the model under consideration under the condition (\ref{rstr-t}).
Direct calculation of relative acceleration
$$
Q\equiv\ddot{a}/a = H^2+\dot{H}
$$
gives
\begin{equation}
Q=\frac{\ddot{a}}{a}=\mu_1^2e^{-2\mu_1 \phi}(1-2\mu_2^3)-2\mu_1 \mu_3e^{-\mu_2\phi}+\mu_3
\end{equation}

Generally speaking the case when $(1-2\mu_2^3)>0$ gives two accelerating periods in terms of $e^{-\mu_2\phi}$ when $\mu_3<0$ and when $0<\mu_3<1/\sqrt{2}$. If $(1-2\mu_2^3)<0$ then there are exist one accelerating period in the cases: i) $\mu_3>1/\sqrt{2}$ solution oscillate between two roots; ii) $0<\mu_3<1/\sqrt{2}$ solution oscillate between zero and the bigger root; iii) the same behaviour when $\mu_3<-1/\sqrt{2}$.

Also, we note that a numerical analysis of the stages of expansion of the universe for the Hubble parameter (\ref{HubbleSR-1}) was performed in \cite{GSTG}.

We also give an asymptotic analysis of the dynamics of expansion of the universe for the case of positive values of the constant parameters $\mu_{1}$, $\mu_{3}$ and $c$.

At the small times that correspond to inflationary stage $t\approx0$ one has the exponential
expansion with following Hubble parameter and scale factor
\begin{equation}
H_{inf}\simeq\frac{\mu_{1}}{c}+\mu_{3},~~~~~~
a_{inf}(t)\propto \exp\left[\left(\frac{\mu_{1}}{c}+\mu_{3}\right)t\right].
\end{equation}

On the following stage, under the condition
\begin{equation}
\frac{\mu_{1}}{2\mu_{1}\mu^{2}_{2}t+c}\gg\mu_{3},
\end{equation}
one has the power-law expansion with the Hubble parameter and scale factor
\begin{equation}
H_{PL}(t)\simeq\frac{\mu_{1}}{2\mu_{1}\mu^{2}_{2}t+c},~~~~~
a_{PL}(t)\propto(2\mu_{1}\mu^{2}_{2}t+c)^{1/2\mu^{2}_{2}}\,.
\end{equation}

At the large times $t\rightarrow\infty$ we have the second accelerated exponential expansion of the universe with
\begin{equation}
H_{sec}\simeq\mu_{3},~~~~~
a_{sec}(t)\propto\exp\left(\mu_{3}t\right)\,.
\end{equation}

We also note that the rate of expansion of the universe during the second inflation is much lower than in the case of the first inflation $H_{sec}\ll H_{inf}$, i.e. one has the condition $\frac{\mu_{1}}{c}\gg\mu_{3}$.

Therefore, the models under consideration implies the exit from the first inflationary accelerated expansion
stage (for $\mu_{2}=\pm1$ we have the dynamics corresponding to the radiation domination stage)
and the second accelerated expansion of the universe as well.
Thus, this dynamics can be considered as the combination of de Sitter and Friedmann solutions and
 corresponds to the correct change in the stages of the universe's expansion.

\section{The exact solutions for exponential power-law inflation}\label{section4}

In this section we study the exact solutions of the equations (\ref{ISB1})--(\ref{ISB2}) for the Hubble parameter (\ref{Hubble}).
These solutions can be noted as
\begin{eqnarray}
\label{POTEPL}
&&V(\phi)=\mu^{2}_{1}(3-2\mu^{2}_{2})\exp(-2\mu_{2}\phi)-6\mu_{1}\mu_{3}\exp(-\mu_{2}\phi)+3\mu^{2}_{3},\\
\label{field}
&&\phi(t)=\frac{1}{\mu_{2}}\ln\left(-2\mu_{1}\mu^{2}_{2}t-c\right),\\
\label{Hubble-2}
&&H(t)=\frac{\mu_{1}}{2\mu_{1}\mu^{2}_{2}t+c}+\mu_{3},\\
\label{scalefactor}
&&a(t)=a_{0}\exp(\mu_{3}t)(2\mu_{1}\mu^{2}_{2}t+c)^{1/2\mu^{2}_{2}},
\end{eqnarray}

As one can see, the evolution of the scalar field and expansion of the universe are the same as for the case of slow-roll approximation.

For the potential (\ref{POTEPLSR-1}) with $\mu_{2}=\sqrt{2/3}$ from the equations (\ref{R})--(\ref{F}) we obtain
\begin{equation}
\label{FRgen}
f(R)=\frac{\mu_{1}}{\mu_{3}}R+\frac{1}{24\mu^{2}_{3}}R^{2}+\frac{8}{3}\mu^{2}_{1},
\end{equation}

If we additionally chose $\mu_{1}=0$ we get de Sitter solution
\begin{equation}
V(\phi)=3\mu^{2}_{3},~~~\phi(t)=\sqrt{\frac{3}{2}}\ln(-c),~~~H(t)=\mu_{3},~~~a(t)=a_{0}\exp(\mu_{3}t),
\end{equation}
which correspond to quadratic term only
\begin{equation}
f_{dS}(R)=\frac{1}{24\mu^{2}_{3}}R^{2}.
\end{equation}

Therefore, the quadratic correction in curvature $R^{2}$ determines the accelerated expansion of the universe in the Starobinsky inflationary model what is correct in the case of the slow-roll approximation analysis as well.

For the case $\mu_{1}=\mu_{3}=\frac{1}{2}m$, from (\ref{FRgen}) and (\ref{POTEPL}) we have the Starobinsky gravity with the additional cosmological constant
\begin{equation}
f(R)=R+\frac{1}{6m^{2}}R^{2}+\frac{2}{3}m^{2},
\end{equation}
and the following potential in Einstein frame
\begin{equation}
\label{Starobinsky_exact}
V(\phi)=\frac{27}{20}m^{2}\left(1-\frac{5}{9}e^{-\sqrt{\frac{2}{3}}\phi}\right)^{2}-\frac{3}{5}m^{2},
\end{equation}
with the same corresponding expressions for evolution of the scalar field and universe expansion (\ref{fieldsr})--(\ref{scalefactorsr}).

Supplementing the model with the positive cosmological constant $\tilde{V}(\phi)\rightarrow V(\phi)+\Lambda$ associated with the nonzero vacuum energy, where $\Lambda=\frac{1}{3}m^{2}$, we obtain $\tilde{V}(\phi=0)=V(\phi=0)+\Lambda=0$, i.e., the true vacuum state for zero scalar field.

Also, we note, that the potential in the case of the exact solutions can be expressed as
\begin{equation}
V(\phi)=V_{SR}(\phi)-2H'^{2}_{\phi},
\end{equation}
and for the case of slow-roll approximation $H'^{2}_{\phi}\approx0$ one has $V(\phi)=V_{SR}(\phi)$.

Thus, the Starobinsky inflation can be considered as the partial case of exponential power-law inflation with the specific choice of the parameters $\mu_{1}, \mu_{2}, \mu_{3}$. For the constant parameter $\mu_{2}\neq\sqrt{2/3}$ one hasn't the explicit expression for the type of $f(R)$-gravity from the equations  (\ref{R})--(\ref{F}). However, one can consider exponential power-law inflation in Einstein frame which is connected with $f(R)$-gravity expressed in parametric form.

\section{The correspondence to the observational constraints}\label{section5}

One of the main methods of verification of cosmological models is the comparison of the obtained parameters of cosmological
perturbations with observational constraints which are based on measurements of CMB anisotropy.

These constraints from the PLANCK observations at the moment estimated as~\cite{Aghanim:2018eyx}
\begin{eqnarray}
\label{PLANCK1}
&&{\mathcal P}_{S}=2.1\times10^{-9},~~~~~~~~~~~~~~~~~n_{S}=0.9663\pm 0.0041,\\
\label{PLANCK2}
&&r<0.1~~~\text{(PLANCK 2018)},~~~~r<0.065~~~\text{(PLANCK 2018/BICEP2/Keck-Array)}.
\end{eqnarray}

The parameters of the cosmological perturbations for Starobinsky and non-minimal Higgs inflation were calculated, for example, in the paper \cite{Mishra:2018dtg}. The spectral index of scalar perturbations $n_{S}$ and tensor-to-scalar ratio $r$ for this type of inflation are $n_{S}=0.967$ and $r=0.003$.

For the exponential power-law inflation with arbitrary constants $\mu_{1}$, $ \mu_{2}$ , $\mu_{3}$ and $c$ these parameters were calculated in the paper \cite{Fomin:2019yls} on the basis of the exact solutions of background dynamic equations. The resulting relation between spectral index of scalar perturbations and tensor-to-scalar ratio was determined as
\begin{equation}
\label{rns}
r=\frac{4s}{n_{S}-3}\left[n_{S}-1+\frac{\mu_{2}\left(\sqrt{2}\mu_{2}
-\sqrt{2\mu^{2}_{2}+4n^{2}_{S}-16n_{S}+12}\,\right)}{\sqrt{2}(n_{S}-3)}\right],
\end{equation}
where the constant parameter $s$ characterizes the normalization of the amplitude of the tensor perturbations.
We will consider the value of this parameter as $s=1$. The value of the power spectrum of the scalar perturbations on the crossing of the Hubble radius ${\mathcal P}_{S}={\mathcal A}^{2}_{S}=2.1\times10^{-9}$ can be always obtained by the choice of the constants $\mu_{1}$, $\mu_{3}$ and $c$ for any $\mu_{2}$ \cite{Fomin:2019yls}.

\begin{figure}[ht]\label{r-nS}
\begin{center}
{\includegraphics*[scale=0.75]{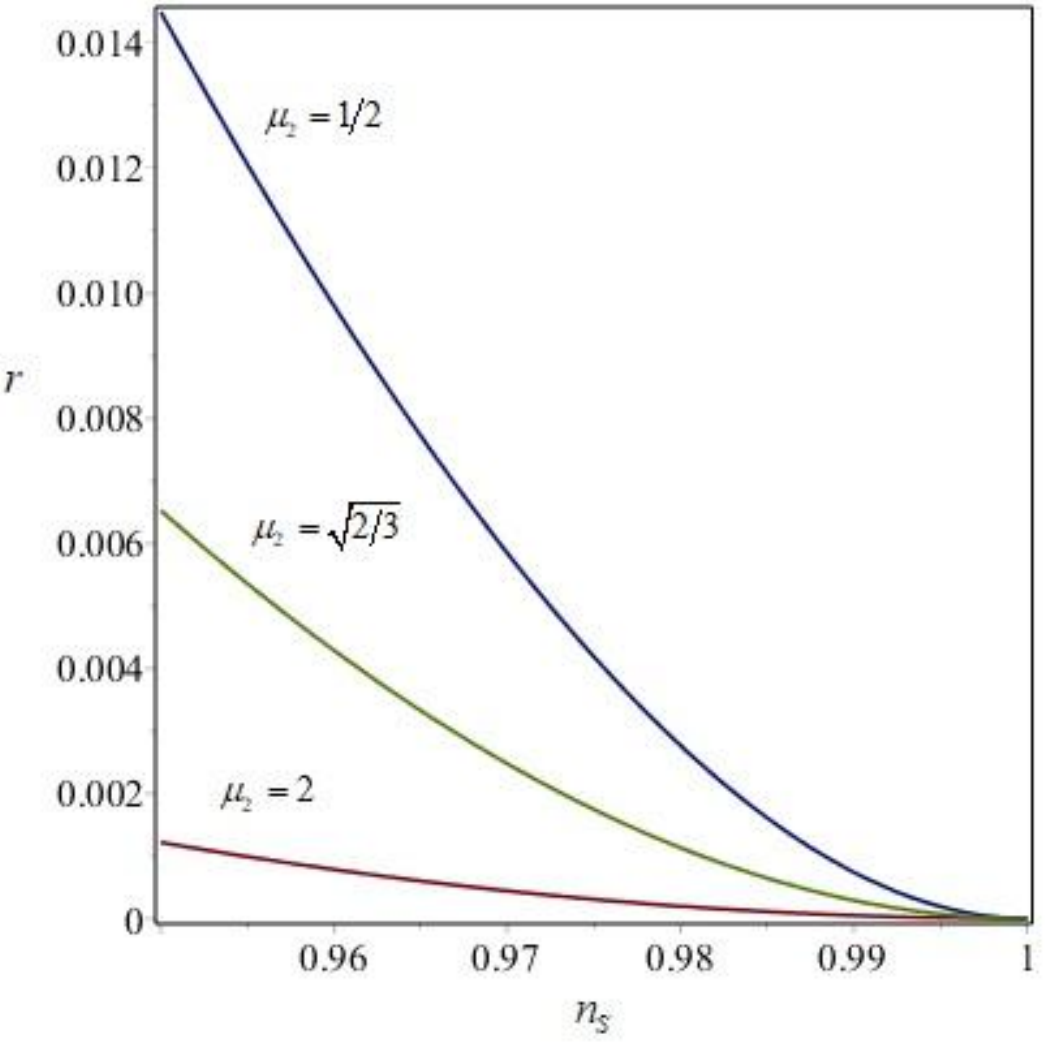}}
\end{center}
\caption{The dependence $r=r(n_{S})$ for different values of the constant $\mu_{2}=1/2,\sqrt{2/3},2$.}
\label{fig1}
\end{figure}

For the case of the Harrison-Zeldovich spectrum ($n_{S}=1$) the expression  (\ref{rns}) gives $r=0$, i.e. the absence of relic gravitational waves (tensor perturbations) for any value of $\mu_{2}$. Also, for $\mu_{2}\rightarrow\infty$ one has $r=0$ for any value of $n_{S}$ as well.

For Starobinsky and non-minimal Higgs inflation with $\mu_{2}=\sqrt{2/3}$ and $n_{S}=0.967$ from the relation (\ref{rns}) one has $r=0.003$ which corresponds to the result obtained in \cite{Mishra:2018dtg}. Also, one can calculate the parameters of cosmological perturbations for the others values of the constant $\mu_{2}$.

On Fig.\ref{fig1} the dependences of the tensor-to-scalar ratio $r$ on the spectral index of scalar perturbations $n_{S}$ for various values of the parameter $\mu_{2}$ which corresponds to the observational constraints are shown.

As one can see, exponential power-law inflation allows to satisfy any restrictions on the value of the tensor-to-scalar ratio by choosing the parameter $\mu_{2}$.

\section{Conclusion}

We consider the exponential power-law inflation on the basis of the exact and approximate cosmological solutions.
The law of evolution of a scalar field, its potential and the nature of the dynamics of the early universe were obtained.
Based on the results obtained, it can be argued that the Starobinsky and non-minimal Higgs models
are the partial cases of the exponential power-law inflation from a dynamically motivated point of view.

On the basis of exact solutions of the equations of cosmological dynamics in Einstein frame an expression for Starobinsky gravity with an additional term corresponding to the cosmological constant
\begin{equation}
f(R)=R+\frac{1}{6m^{2}}R^{2}+\frac{2}{3}m^{2},
\end{equation}
was obtained.
Also, it was shown that the pure exponential expansion of the early universe associated with quadratic correction in curvature $R^{2}$.

An analysis of the observational constraints on the values of the parameters of cosmological perturbations suggests that exponential power-law inflation can satisfy any constraints on the value of the tensor-to-scalar ratio obtained from the observational data.

\section{Acknowledgements}

I.V. Fomin and S.V. Chervon were supported by RFBR Grant 18-52-45016 IND a.
S.V.C. is grateful for support by the Program of Competitive Growth of Kazan Federal University.

%%%%====================bibliography=================================

\end{document}